\input harvmac.tex
\input epsf.tex
\def\figin{\epsfcheck\figin}\def\figins{\epsfcheck\figins}
\def\epsfcheck{\ifx\epsfbox\UnDeFiNeD
\message{(NO epsf.tex, FIGURES WILL BE IGNORED)}
\gdef\figin##1{\vskip2in}\gdef\figins##1{\hskip.5in}
\else\message{(FIGURES WILL BE INCLUDED)}%
\gdef\figin##1{##1}\gdef\figins##1{##1}\fi}
\def\DefWarn#1{}
\def\figinsert{\goodbreak\midinsert}
\def\ifig#1#2#3{\DefWarn#1\xdef#1{fig.~\the\figno}
\writedef{#1\leftbracket fig.\noexpand~\the\figno}%
\figinsert\figin{\centerline{#3}}\medskip\centerline{\vbox{\baselineskip12pt
\advance\hsize by -1truein\noindent\footnotefont{\bf Fig.~\the\figno:} #2}}
\bigskip\endinsert\global\advance\figno by1}


\lref\review{
O.~Aharony, S.~S.~Gubser, J.~Maldacena, H.~Ooguri and Y.~Oz,
``Large N field theories, string theory and gravity,''
Phys.\ Rept.\  {\bf 323}, 183 (2000)
[hep-th/9905111].
}

\lref\vijaytrivedi{
V.~Balasubramanian, P.~Kraus, A.~E.~Lawrence and S.~P.~Trivedi,
Phys.\ Rev.\ D {\bf 59}, 104021 (1999)
[hep-th/9808017].
}

\lref\multiple{
S.~Aminneborg, I.~Bengtsson, D.~Brill, S.~Holst and P.~Peldan,
``Black holes and wormholes in 2+1 dimensions,''
Class.\ Quant.\ Grav.\  {\bf 15}, 627 (1998)
[gr-qc/9707036];
D.~Brill,
``Black holes and wormholes in 2+1 dimensions,''
gr-qc/9904083.
}

\lref\swsingular{
N.~Seiberg and E.~Witten,
``The D1/D5 system and singular CFT,''
JHEP {\bf 9904}, 017 (1999)
[hep-th/9903224].
}

\lref\krasnov{K.~Krasnov,
``Holography and Riemann surfaces,''
hep-th/0005106.
}

\lref\bousso{
R.~Bousso,
``A Covariant Entropy Conjecture,''
JHEP {\bf 9907}, 004 (1999)
[hep-th/9905177];
R.~Bousso,
JHEP {\bf 9906}, 028 (1999)
[hep-th/9906022].
}

\lref\wittenyau{
E.~Witten and S.~T.~Yau,
``Connectedness of the boundary in the AdS/CFT correspondence,''
Adv.\ Theor.\ Math.\ Phys.\  {\bf 3}, 1635 (1999)
[hep-th/9910245].
}
\lref\precusors{
J.~Polchinski, L.~Susskind and N.~Toumbas,
``Negative energy, superluminosity and holography,''
Phys.\ Rev.\ D {\bf 60}, 084006 (1999)
[hep-th/9903228].
}

\lref\teschner{
J.~Teschner,
``On structure constants and fusion rules in the SL(2,C)/SU(2) WZNW  model,''
Nucl.\ Phys.\ B {\bf 546}, 390 (1999)
[hep-th/9712256].
}

\lref\zam{L. Fateev, A. Zamolodchikov and Al. Zamolodchikov, unpublished }

\lref\ssug{
L.~Susskind and J.~Uglum,
``Black hole entropy in canonical quantum gravity and superstring theory,''
Phys.\ Rev.\ D {\bf 50}, 2700 (1994)
[hep-th/9401070].
}

\lref\cwil{
C.~Callan and F.~Wilczek,
``On geometric entropy,''
Phys.\ Lett.\ B {\bf 333}, 55 (1994)
[hep-th/9401072];
C.~Holzhey, F.~Larsen and F.~Wilczek,
``Geometric and renormalized entropy in conformal field theory,''
Nucl.\ Phys.\ B {\bf 424}, 443 (1994)
[hep-th/9403108].
}

\lref\ss{
M.~Spradlin and A.~Strominger,
``Vacuum states for AdS(2) black holes,''
JHEP {\bf 9911}, 021 (1999)
[hep-th/9904143].
}

\lref\hp{
S.~W.~Hawking and D.~N.~Page,
``Thermodynamics Of Black Holes In Anti-De Sitter Space,''
Commun.\ Math.\ Phys.\  {\bf 87}, 577 (1983).
}

\lref\hm{
G.~T.~Horowitz and D.~Marolf,
``A new approach to string cosmology,''
JHEP {\bf 9807}, 014 (1998)
[hep-th/9805207].
}
\lref\gilad{
G.~Lifschytz and M.~Ortiz,
``Scalar field quantization on the (2+1)-dimensional black hole background,''
Phys.\ Rev.\ D {\bf 49}, 1929 (1994)
[gr-qc/9310008].
}
\lref\esko{
E.~Keski-Vakkuri,
``Bulk and boundary dynamics in BTZ black holes,''
Phys.\ Rev.\ D {\bf 59}, 104001 (1999)
[hep-th/9808037].
}
\lref\steif{
A.~R.~Steif,
``The Quantum stress tensor in the three-dimensional black hole,''
Phys.\ Rev.\ D {\bf 49}, 585 (1994)
[gr-qc/9308032].
}
\lref\shiraishi{
K.~Shiraishi and T.~Maki,
``Vacuum polarization around a three-dimensional black hole,''
Class.\ Quant.\ Grav.\  {\bf 11}, 695 (1994).
}

\lref\lowe{
D.~A.~Lowe and L.~Thorlacius,
``AdS/CFT and the information paradox,''
Phys.\ Rev.\ D {\bf 60}, 104012 (1999)
[hep-th/9903237].
}

\lref\vijay{
V.~Balasubramanian, P.~Kraus and A.~E.~Lawrence,
``Bulk vs. boundary dynamics in anti-de Sitter spacetime,''
Phys.\ Rev.\ D {\bf 59}, 046003 (1999)
[hep-th/9805171].
}

\lref\hh{
J.~B.~Hartle and S.~W.~Hawking,
``Path Integral Derivation Of Black Hole Radiance,''
Phys.\ Rev.\ D {\bf 13}, 2188 (1976).
}

\lref\ms{
J.~Maldacena and A.~Strominger,
``AdS(3) black holes and a stringy exclusion principle,''
JHEP {\bf 9812}, 005 (1998)
[hep-th/9804085].
}

\lref\msh{
P. Martin and J. Schwinger, Phys. Rev. {\bf 115}, 1342 (1959);
J. Schwinger, JMP vol 2, 407(1961);
 K.T. Mahanthappa, Phys.Rev.vol 126, 329(1962);
P.M. Bakshi and K.T. Mahanthappa, JMP vol 4, 1 and 12 (1963);
Keldysh  JETP Vol 47,  
1515(1964);
Y. Takahashi and H. Umezawa, 
Collective Phenomena {\bf 2} 55 (1975). }

\lref\israel{W.~Israel,
``Thermo Field Dynamics Of Black Holes,''
Phys.\ Lett.\ A {\bf 57}, 107 (1976).
}

\lref\mss{
J.~M.~Maldacena and L.~Susskind,
``D-branes and Fat Black Holes,''
Nucl.\ Phys.\ B {\bf 475}, 679 (1996)
[hep-th/9604042].
}

\lref\curvedqft{
 N. Birrel and P. Davies, ``Quantum Fields in curved 
space'' Cambridge Univ. Press, (1982) ;
S.~A.~Fulling, 
``Aspects Of Quantum Field Theory In Curved Space-Time,''
 Cambridge Univ. Press, (1989) ;
R.~M.~Wald,
``Quantum field theory in curved space-time and black hole thermodynamics,''
 Chicago  Univ. Press (1994).
}

\lref\btzpaper{M.~Banados, C.~Teitelboim and J.~Zanelli,
``The Black hole in three-dimensional space-time,''
Phys.\ Rev.\ Lett.\  {\bf 69}, 1849 (1992)
[hep-th/9204099];
M.~Banados, M.~Henneaux, C.~Teitelboim and J.~Zanelli,
``Geometry of the (2+1) black hole,''
Phys.\ Rev.\ D {\bf 48}, 1506 (1993)
[gr-qc/9302012].
}

\lref\unruh{
W.~G.~Unruh,
``Notes On Black Hole Evaporation,''
Phys.\ Rev.\ D {\bf 14}, 870 (1976).
}

\lref\dmmv{
R.~Dijkgraaf, J.~Maldacena, G.~Moore and E.~Verlinde,
``A black hole farey tail,''
hep-th/0005003.
}
\lref\polchinski{
J.~Polchinski,
``S-matrices from AdS spacetime,''
hep-th/9901076.
}

\lref\hawkinginfo{
S.~W.~Hawking,
``Breakdown Of Predictability In Gravitational Collapse,''
Phys.\ Rev.\ D {\bf 14}, 2460 (1976).
}

\lref\thooft{
G.~'t Hooft,
``Ambiguity Of The Equivalence Principle And Hawking's Temperature,''
J.\ Geom.\ Phys.\  {\bf 1}, 45 (1984).
}

\lref\wittenthermal{
E.~Witten,
``Anti-de Sitter space, thermal phase transition, and confinement in  gauge theories,''
Adv.\ Theor.\ Math.\ Phys.\  {\bf 2}, 505 (1998)
[hep-th/9803131].
}

\lref\wittenone{
E.~Witten,
``Anti-de Sitter space and holography,''
Adv.\ Theor.\ Math.\ Phys.\  {\bf 2}, 253 (1998)
[hep-th/9802150].
}

\lref\gibhaw{
G.~W.~Gibbons and S.~W.~Hawking,
``Action Integrals And Partition Functions In Quantum Gravity,''
Phys.\ Rev.\ D {\bf 15}, 2752 (1977).
}

\lref\kostas{
M.~Henningson and K.~Skenderis,
``The holographic Weyl anomaly,''
JHEP {\bf 9807}, 023 (1998)
[hep-th/9806087];
S.~de Haro, K.~Skenderis and S.~N.~Solodukhin,
``Gravity in warped compactifications and 
the holographic stress tensor,''
hep-th/0011230.
}
\lref\carnero{
B.~Carneiro da Cunha,
``Inflation and holography in string theory,''
hep-th/0105219.
}

\lref\gkp{
S.~S.~Gubser, I.~R.~Klebanov and A.~M.~Polyakov,
``Gauge theory correlators from non-critical string theory,''
Phys.\ Lett.\ B {\bf 428}, 105 (1998)
[hep-th/9802109].
}

\lref\jmadscft{
J.~Maldacena,
``The large N limit of superconformal field theories and supergravity,''
Adv.\ Theor.\ Math.\ Phys.\  {\bf 2}, 231 (1998)
[Int.\ J.\ Theor.\ Phys.\  {\bf 38}, 1113 (1998)]
[hep-th/9711200].
}

\lref\jacobson{
T.~A.~Jacobson,
``Introduction to Black Hole Microscopy,''
hep-th/9510026; 
T.~Jacobson,
``A Note on Hartle-Hawking vacua,''
Phys.\ Rev.\ D {\bf 50}, 6031 (1994)
[gr-qc/9407022].
}

\lref\JacobsonMI{
T.~Jacobson,
``On the nature of black hole entropy,''
gr-qc/9908031.
}

\lref\gary{G.~T.~Horowitz and D.~L.~Welch,
``Exact three-dimensional black holes in string theory,''
Phys.\ Rev.\ Lett.\  {\bf 71}, 328 (1993)
[hep-th/9302126].
}

\lref\geonads{
J.~Louko and D.~Marolf,
``Single-exterior black holes and the AdS-CFT conjecture,''
Phys.\ Rev.\ D {\bf 59}, 066002 (1999)
[hep-th/9808081];
J.~Louko,
``Single-exterior black holes,''
Lect.\ Notes Phys.\  {\bf 541}, 188 (2000)
[gr-qc/9906031];
J.~Louko, D.~Marolf and S.~F.~Ross,
``On geodesic propagators and black hole holography,''
Phys.\ Rev.\ D {\bf 62}, 044041 (2000)
[hep-th/0002111].
}

\lref\complementarity{
G.~'t Hooft,
``The Black Hole Interpretation Of String Theory,''
Nucl.\ Phys.\ B {\bf 335}, 138 (1990);
L.~Susskind, L.~Thorlacius and J.~Uglum,
``The Stretched horizon and black hole complementarity,''
Phys.\ Rev.\ D {\bf 48}, 3743 (1993)
[hep-th/9306069].
}

\lref\joeandy{
J.~Polchinski and A.~Strominger,
``A Possible resolution of the black hole information puzzle,''
Phys.\ Rev.\ D {\bf 50}, 7403 (1994)
[hep-th/9407008].
}

\lref\andyhouches{
A.~Strominger,
``Les Houches lectures on black holes,''
hep-th/9501071.
}

\lref\hw{
P.~Horava and E.~Witten,
``Eleven-Dimensional Supergravity on a Manifold with Boundary,''
Nucl.\ Phys.\ B {\bf 475}, 94 (1996)
[hep-th/9603142].
}

\lref\karch{
A.~Karch and L.~Randall,
``Open and closed string interpretation of 
SUSY CFT's on branes with  boundaries,''
JHEP {\bf 0106}, 063 (2001)
[hep-th/0105132].
}
\lref\hfa{
M.~Fabinger and P.~Horava,
``Casimir effect between world-branes in heterotic M-theory,''
Nucl.\ Phys.\ B {\bf 580}, 243 (2000)
[hep-th/0002073].
}

\lref\son{
 J.~Maldacena, H.~Ooguri and J.~Son,
``Strings in AdS(3) and the SL(2,R) WZW model. II: 
Euclidean black hole,''
hep-th/0005183.
}

\lref\gallowaytwo{
G.~J.~Galloway, K.~Schleich, D.~Witt and E.~Woolgar,
``The AdS/CFT correspondence conjecture and topological censorship,''
Phys.\ Lett.\ B {\bf 505}, 255 (2001)
[arXiv:hep-th/9912119].
}

\lref\gallowayone{
G.~J.~Galloway, K.~Schleich, D.~M.~Witt and E.~Woolgar,
``Topological censorship and higher genus black holes,''
Phys.\ Rev.\ D {\bf 60}, 104039 (1999)
[arXiv:gr-qc/9902061].
}

\lref\horava{
P.~Horava,
``Background Duality Of Open String Models,''
Phys.\ Lett.\ B {\bf 231}, 251 (1989).
}

\lref\jacobsonthree{
T.~Jacobson,
``On the nature of black hole entropy,''
arXiv:gr-qc/9908031.
}


\Title{\vbox{
\hbox{NSF-ITP-01-59}
\hbox{\tt hep-th/0106112}}}{
Eternal black holes in Anti-de-Sitter 
}
\bigskip
\centerline{ Juan Maldacena }
\bigskip
\centerline{ Jefferson Physical Laboratory}
\centerline{Harvard University}
\centerline{Cambridge, MA 02138, USA}
\bigskip
\centerline{ Institute for Advanced Study}
\centerline{Princeton, NJ 08540}

\vskip .3in

We propose a dual non-perturbative description for  maximally 
extended Schwarzschild Anti-de-Sitter spacetimes.
The description involves 
two copies of the conformal field theory associated to the AdS spacetime
and an  initial entangled state. In this context  we 
also discuss a version of the  information loss
paradox  and its resolution.

\bigskip

\vskip 3in

\newsec{ Introduction}

\ifig\penrose{Penrose diagram of the extended AdS Schwarzschild geometry.
Region I covers the region that is outside the horizon from the point
of view of an observer on the right boundary. Region II is an identical
copy and includes a second boundary. Regions III and IV contain
spacelike singularities. The diagram shows the time and radial directions, 
over each point there is a sphere $S^{d-1}$. This sphere shrinks as we 
approach the 
singularities.  }
{\epsfxsize1.5in\epsfbox{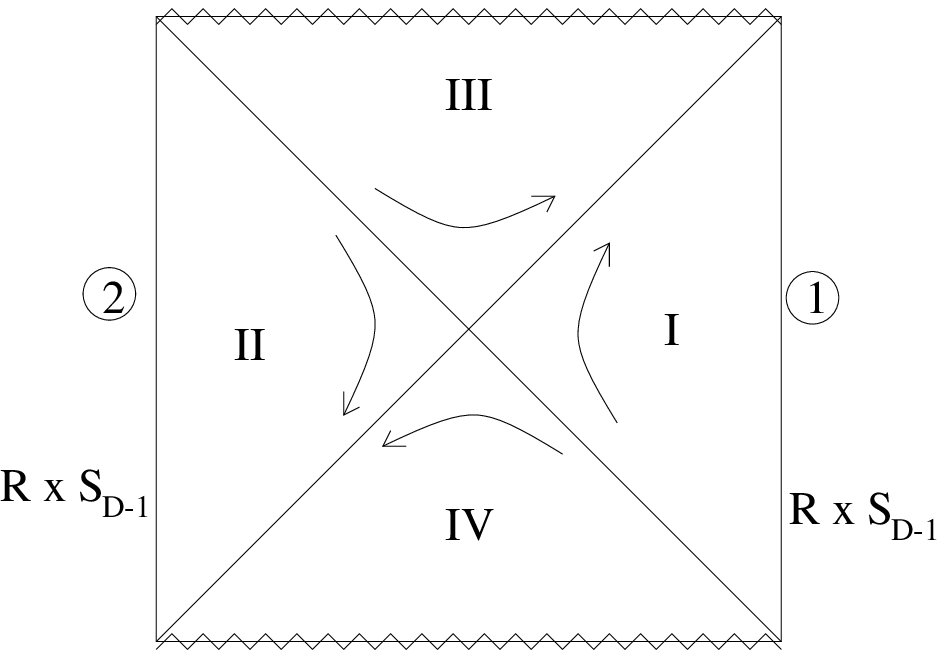}}

 An eternal black hole has an extended Penrose diagram which is 
depicted in  \penrose .  
 This Penrose diagram has two asymptotically 
AdS regions. From the point of view of each of these regions the other
region is behind the horizon. It   is a  time 
dependent spacetime  since there is no global  timelike isometry. 
The regions close to the spacelike singularities can be viewed 
as big-bang or big-crunch cosmologies (which are homogeneous but
not isotropic). 
 We will  propose that this spacetime
can be holographically described by considering two identical, 
non-interacting copies of the conformal field theory and picking a 
particular entangled state. This point of view is based on 
Israel's description of eternal black holes \israel . 
 A similar observation in the context of AdS/CFT  was made in 
\refs{\hm,\vijaytrivedi,\carnero} \foot{In \vijaytrivedi\  the
formula for the entangled state as a function 
of the temperature  is off by a  factor of 2.
It seems to be the related to the  factor of 2 that led to the claim 
\thooft\  that
the black hole temperature is twice what Hawking originally computed.}.  
Here we will emphasize that by including both copies
we naturally get a description of the interior region of black holes, 
including the region near the singularities.
This holographic description can be viewed as a resolution of the 
initial and final singularities.

Using this correspondence we can study some aspects of the information 
loss paradox. 
 We will formulate a precise calculation
on  the eternal black hole spacetime of  \penrose . The
result of this calculation  shows information loss. We will 
show that  information can be  preserved after summing over geometries.

\newsec{The correspondence}

We  start with an $AdS_{d+1}$ spacetime  and its holographic dual 
conformal field theory CFT$_d$, as in \refs{\jmadscft,\gkp,\wittenone}
(for a review see \review ).
The conformal field theory is defined on a cylinder $R\times S^{d-1}$.
This cylinder is also the boundary of AdS. 
A general conclusion of the studies of AdS/CFT is that the boundary 
conditions in AdS specify the theory and the  normalizable
modes in the interior correspond to  states \refs{\vijay,\review}.
 When we give a 
particular spacetime which is asymptotically AdS we are giving a CFT and
a particular state in the CFT.  

\ifig\kruskal{The Lorentzian black hole in Kruskal coordinates. 
The singularity is at $uv=1$ and the boundary of $AdS$ is at $uv=-1$. 
}
{\epsfxsize1.5in\epsfbox{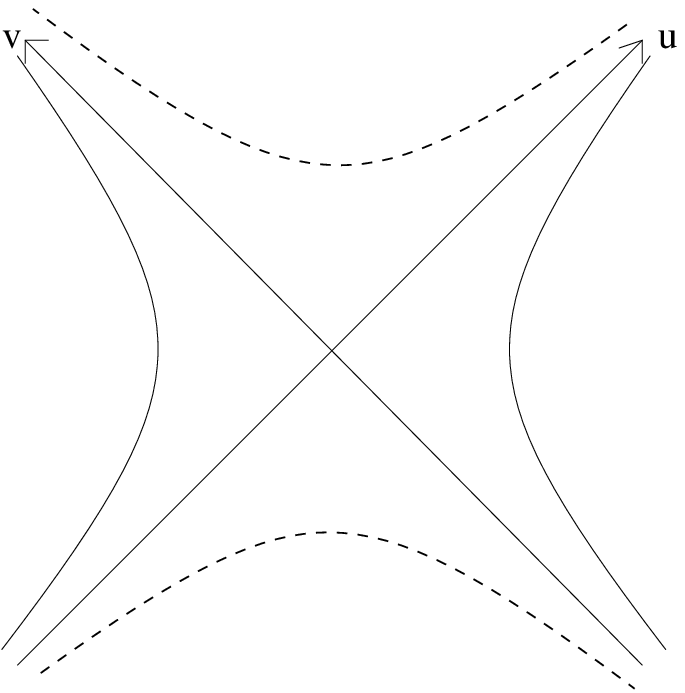}}

We  consider the so called ``big'' black holes
in $AdS$. These are black holes which have positive specific heat and are
the dominant contribution to the 
 canonical thermodynamic   ensemble.
In our analysis below we will write explicit formulas for 
 black holes in   $AdS_3$ since calculations are easiest in that case,
 but everything we say also holds for big black 
holes in 
$AdS_{d+1}$, $d\geq 2$. 
The Euclidean metric of three dimensional 
 black hole can be written in the 
following equivalent forms \btzpaper\ 
\eqn\btzmet{\eqalign{
ds^2  =&   (r^2-1) {d\tilde \tau^2 } + { dr^2 \over r^2 -1}
+ r^2 d \tilde \phi^2 
\cr
ds^2 =& 4 { dz d \bar z \over ( 1 - |z|^2)^2} +  { (1 + |z|^2 )^2 
\over (1 -  |z|^2)^2 } d \tilde \phi^2 
\cr
 \tilde \phi =& {2 \pi \phi  \over  \beta } ~,~~~~~ \tilde \tau = 
{2 \pi \tau  \over  \beta  } ~,~~~~~~~ z = |z|e^{ i \tilde \tau } 
\cr 
\phi =& \phi + 2 \pi ~,~~~~~~ \tau = \tau + \beta 
}}
where we have set $R_{AdS} = 1$. $\tau, \phi$ are to 
be thought of as the coordinates of the space on which the CFT is defined
so that $\beta^{-1} $ is the temperature. Notice that the boundary of AdS
is at $|z|=1$.
By analytically continuing in the imaginary   part of $z$ in \btzmet ,
and
setting $z = -v  , \bar z = u  $,  we 
obtain the eternal black hole in Kruskal coordinates 
\eqn\kruskalcoord{
ds^2 = { - 4 du dv \over (1 + uv)^2 } + {(1 - uv)^2 \over (1 + uv)^2 } 
d \tilde \phi^2 }  
where $u = t + x , v = t - x$. 
In these  coordinates the spacetime looks as in  \kruskal . 
The event horizons are at $u=0$ and $v=0$. The boundary of AdS is
at $uv=-1$ and the past and future singularities are at $uv = 1$. 
In these coordinates it is clear that nothing special happens at the
horizon. 
Notice that the metric \kruskalcoord\ is time dependent. It has 
the boost-like isometry that acts by $ u \to e^\lambda u, ~
v\to   e^{-\lambda}v $. This is the usual ``time'' translation invariance
in Schwarzschild coordinates. The orbits of these isometries are time-like
going forward in time in region I, timelike going backwards in time
in region II and spacelike in regions III and IV, see  \penrose . 
The metric \kruskalcoord\ has a reflection symmetry under $t \to -t$. 
In fact,  we can glue the $Im(z)=0$ cross section of the Euclidean metric
\btzmet , to the $t=0$ spatial cross section of \kruskalcoord . 
We can view the Euclidean part of the geometry as giving the initial
wavefunction which we then evolve in Lorentzian signature. This is 
the Hartle-Hawking construction of the wavefunction \hh .

\ifig\gluing{ The Hartle-Hawking-Israel wavefunction can be 
thought of as arising from gluing half the Euclidean geometry 
at $t=0$ to half the Lorentzian geometry. Over each point on  this
diagram there is a sphere $S^{d-1}$.
}
{\epsfxsize1.5in\epsfbox{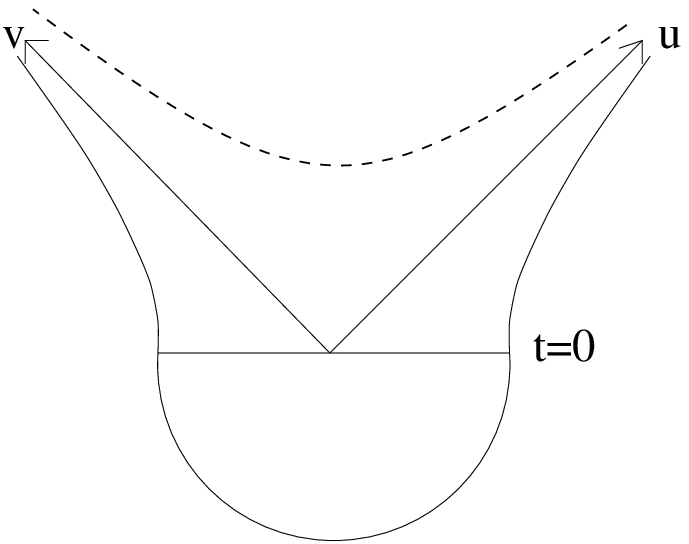}}

Let us understand what this wavefunction is from the point of view
of the boundary CFT. The boundary of the Euclidean black hole 
is $S^1_\beta \times S^{d-1} $. The section of the 
 Euclidean metric
\btzmet\ at $Im(z)=0$ intersects the boundary on two disconnected
spheres $S^{d-1}  + S^{d-1} $. (In the three dimensional case \btzmet\
we get two circles  $S^1_\phi + S^1_\phi$). 
 The Euclidean time direction 
connects these two spheres. The path integral of the boundary CFT 
is then over $I_{\beta/2} \times S^{d-1}$, where $I_{\beta/2}$
is an interval of length $\beta/2$. This path integral 
 then gives a wavefunction on the product of two
copies of the CFT. Let ${\cal H} ={\cal H}_1 \times {\cal H}_2$
denote the full Hilbert space consisting of two copies of the 
Hilbert space of the CFT. The wavefunction
 $|\Psi \rangle \in {\cal H}$
is 
\eqn\wavefunction{
|\Psi \rangle = { 1 \over \sqrt{Z(\beta)} }
 \sum_{n} e^{- \beta E_n/2} | E_n\rangle_1 \times
| E_n\rangle_2 
}
The sum runs over all energy eigenstates  the system and the 
subindex 1,2 indicates the Hilbert space where the state is defined. 
 $Z(\beta)$ is the partition function 
of one copy of the CFT at temperature $\beta^{-1}$ and it is necessary in 
\wavefunction\ for normalization purposes.
We view this wavefunction as an initial condition for Lorentzian evolution. 
More precisely, this is the wavefunction at Lorentzian time 
$t_1=t_2 =0$. We have two times since we have two independent 
copies of the field theory and we can evolve as much as we want in
each time. Note, however, that the state $|\Psi \rangle$ is invariant
under $\tilde H = H_1 - H_2 $. 
This construction with two copies of a field theory in a pure state 
given by \wavefunction\ is very familiar in the description of 
real time thermal field theories and goes under the name of
``thermofield  dynamics'' \msh . 
Thermal expectation values in any  field theory can be calculated 
as conventional quantum mechanical expectation values in two copies
of the field theory in the inital pure state \wavefunction .
In other words 
\eqn\thermo{
\langle \Psi | {\cal O}_1 | \Psi \rangle = Tr[ \rho_\beta {\cal O}_1 ]
}
where ${\cal O}_1$ is any operator defined on the first copy of the
field theory. Since  the left hand side of \thermo\ contains 
no operators acting on the second copy 
of the field theory we can sum over all  states of the second copy
of the field theory and obtain 
the result on the right hand side of \thermo . 
After doing this sum we get the thermal 
density matrix in the first copy of the field theory. 
One views the thermal density matrix as arising from entanglement. 
The entropy is the entanglement entropy \msh .

The proposal is that {\it two copies of the CFT in the particular pure
(entangled) state \wavefunction\  is approximately described by 
gravity on the extended AdS Schwarzschild spacetime }. 
The meaning of the word ``approximately'' will become clear later. 
The ``boost'' symmetry of the AdS-Schwarzschild spacetime is the same
as the symmetry under $\tilde H$ of the two copies of the CFT and the
state \wavefunction . 
If we do not do any observations in the second copy of the conformal 
field theory, i.e. we do not insert any operators on the second boundary
of the AdS Schwarzschild spacetime, then  expectation values on the 
first copy become thermal expectation values. The connection between
the  extended Schwarzschild geometry 
and ``thermofield dynamics'' was first noticed 
by Israel \israel , see also \refs{\unruh,\jacobson}.
Here we are just pointing out that that in the
context of AdS-Schwarzschild geometries this connection 
becomes precise and that it gives, in principle, 
 a way to describe the interior. 
The fact that  the eternal black hole in AdS is related to an entangled 
state in the CFT was  observed in \hm ,  \vijaytrivedi .
The fact that black hole entropy and
entanglement entropy are related 
was observed in \refs{\ssug,\cwil}.

In the semiclassical approximation, besides giving a geometry, we 
need to specify the state of all fields living on this geometry. 
In general time dependent geometries 
there is no obvious way to construct the state. In our case
the state of the field  is fixed by patching
the Euclidean solution as described above which gives the standard
Hartle-Hawking state. 
 One can also specify the state of a quantum field
on a general background by specifying the set of positive energy
wavefunctions, see \curvedqft . 
The Hartle-Hawking state is obtained if one defines this
set to be the set of wavefunctions which  restricted to 
$u=0$ ($v=0$)  have an expansion in terms of $e^{ - i \omega v} $
($e^{-i\omega u}$) with $ \omega >0$. 
This Hartle-Hawking-Israel state is such that the 
expectation value of the stress tensor is non singular  (except at the 
past and future singularities). 

In the $AdS_3$ case it is very easy to construct this state since
the BTZ black hole is a quotient of $AdS_3$. This quotient 
is the one that makes $\phi$ periodic \btzpaper .
 If $\phi$ were non-compact we
would have an infinite non extremal black string, 
whose extended Penrose diagram
is  global AdS. 
It can be seen that the H-H prescription of gluing half the Euclidean
solution coincides with the prescription that gives the global AdS vacuum,
which is to glue in half of an infinite Euclidean cylinder. In 
both cases we glue in half a three ball. 
This implies that the notion of positive
frequency as defined in global AdS coincides with the notion of positive
frequency defined in the Hartle-Hawking 
 vacuum of the infinite black string.\foot{
This is related to the observation in \ms\ this non-extremal 
black string can be understood as the usual CFT on $R^2$ 
 in the usual  Minkowski vacuum
but viewed in Rindler space.} By 
doing the quotient we restrict the set of possible wavefunctions but we
do not change the notion of positive frequencies. This implies that 
we can obtain  Green's functions on the Hartle-Hawking state by 
taking the usual Green's functions on global AdS and adding over 
all the images under the group that generates the quotient. 
For Schwarzschild-AdS black holes in other dimensions
 one would have to work harder but the procedure is 
a straight forward analytic continuation from the Euclidean 
solution.\foot{Vacua for  black holes in  $AdS_2$ were 
discussed in \ss .}

Now that we have specified the state 
we can compute, in the semiclassical
approximation,   the correlation functions 
for insertions of operators on various boundaries.
Let us  consider a scalar field in $AdS_3$ which corresponds to a CFT
operator of dimension $(L_0,\bar L_0) = (\Delta,\Delta)$. By summing
over all images we obtain the time ordered correlator for two 
operators inserted on the same boundary  \esko\ 
\eqn\same{
\langle \Psi| T( {\cal O}_1 (t, \phi) {\cal O}_1 (0, 0)) |\Psi \rangle 
\sim  \sum_{n=-\infty}^\infty { 1 \over
\left[ 
\cosh ({2 \pi t  \over \beta}) - 
\cosh ({2 \pi (\phi + 2 \pi n) \over \beta}) - i \epsilon)
\right]^{2 \Delta}  }
 }
For operators inserted on opposite boundaries we obtain
\eqn\other{\langle\Psi| {\cal O}_1 (t_1, \phi_1) {\cal O}_2 (t_2, \phi_2) 
|\Psi \rangle 
\sim  \sum_{n=-\infty}^\infty { 1 \over
\left( 
\cosh ({2 \pi (t_1 + t_2)  \over \beta}) + 
\cosh ({2 \pi (\phi_1-\phi_2 + 2 \pi n) \over \beta})
\right)^{2 \Delta}  }
}
where  we have defined
time on the second copy of the CFT so that it increases towards the
future.
The reason that we get a non-vanishing correlator despite the fact that
the operators live in two decoupled field theories is due to the
fact that we have an entangled state, as in  the  EPR experiment. 
Operators on different boundaries commute.

\ifig\particles{ We can add particles to the
 Hartle-Hawking-Israel state by acting with operators on the 
Euclidean field theory. 
}
{\epsfxsize1.5in\epsfbox{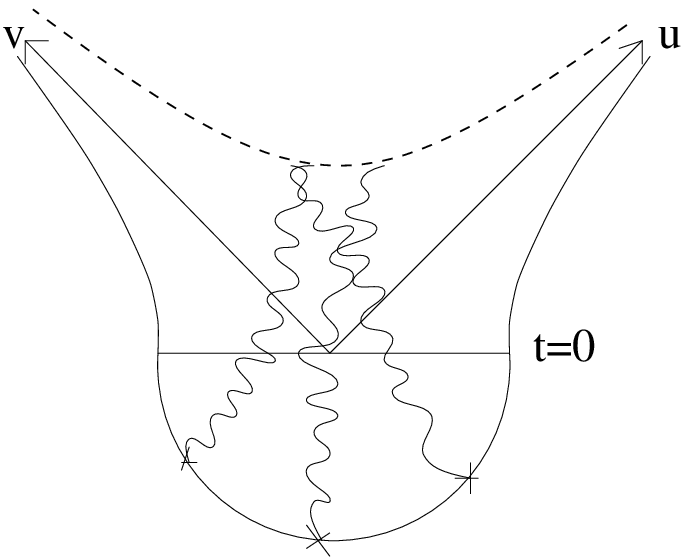}}

It is easy to see how to insert particles in the interior of the 
extended AdS-Schwarzschild spacetime. 
A particle is a small fluctuation around some state, it is a small
deformation of the wavefunction. The H-H wavefunction is defined
by doing the path integral over half the Euclidean geometry. 
We can generate the H-H state plus
some particles  by gluing the same Euclidean geometry but
now with some operators inserted at the boundary, see  \particles . 
This deforms slightly 
the wavefunction and we will not have the state \wavefunction\ but
a slightly different one. 
In the case of $AdS_3$ we can do this  explicitly and say  that an 
operator inserted at a  point ($z_0= e^{i\theta_0},\phi_0$)  along 
the Euclidean boundary (with $- \pi \leq \theta_0 \leq 0$ )
  creates a particle with the Lorentzian 
 wavefunction 
\eqn\lorwave{
\phi(t,x,\phi) \sim { (1 +uv )^{2 \Delta} \over
\left\{ ( 1 -uv) [\cosh({2 \pi (\phi- \phi_0 + 2 \pi n)\over \beta})  -1]
+ (u- e^{-i\theta_0})(v+e^{i\theta_0})
 \right\}^{2 \Delta} }
}
 It is easy to see that this wavefunction
has positive frequency, in the Hartle-Hawking sense.
 We could form other
wavefunctions by convoluting the operator with appropriate functions 
of the boundary point. 
In  principle we can also insert particles by acting with 
operators in the far past in the Lorentzian description. From the 
CFT point of view this is clear. The fact that we can create particles
in region IV (see  \penrose ) by insertion of boundary operators
is essentially   the idea of complementarity 
\complementarity .

Note that these particles are created by operators that are acting
on both copies of the field theory. As a simple example, let us 
replace the CFT by a single harmonic oscillator. The state 
\wavefunction\ becomes 
\eqn\harmosc{
|\psi \rangle = {1 \over \sqrt{Z} } e^{e^{-\beta w/2}
a_1^\dagger a_2^\dagger } |0\rangle
} 
 This state is related by 
a Bogoliubov transformation to the vacuum so that it looks like
the vacuum for the oscillators 
\eqn\newosc{\eqalign{
\tilde a_1^\dagger = & \cosh \theta 
a_1^\dagger - \sinh\theta a_2 ~,~~~~~~~
\tilde a_2^\dagger =  \cosh \theta 
a_2^\dagger - \sinh\theta a_1 
\cr
\tilde a_i |\psi\rangle =&0 ~,~~~~~~~ \tanh \theta = e^{-\beta w/2}
}}
We conclude that particles are created by the 
oscillators $\tilde a_i^\dagger$ and these involve operators on both 
decoupled theories.

It is also easy to see that the expectation value of the stress tensor
for quantum fields in the BTZ  geometry  diverges as we approach the 
singularities \refs{\gilad,\shiraishi}.
 If we define the stress tensor via a point splitting 
procedure, as explained in \curvedqft  , then we will 
need to compute the bulk-bulk 
 two point function for fields in the interior. 
The  two point function can be obtained  by summing over all the images. 
Away from the singularity the images are spacelike separated. As 
we approach the singularity the images become light-like separated
and on the other side they would be timelike separated. The divergence
of the stress tensor is just due to the standard divergence of the
two point function at lightlike separated points.

A similar argument shows that for the rotating black hole in $AdS_3$, 
i.e. the black hole with angular 
momentum along the $\phi$ circle,  the quantum 
stress tensor is singular at the inner horizon \steif .  The 
killing vector associated  to  the identification becomes lightlike at
the singularity. The generator of the identification is obtained by 
exponentiating the action of this Killing vector. 
Once we exponentiate it is possible to 
get images that are lightlike separated before we get to the singularity, 
in fact we get them as soon as we cross the inner horizon. 
This does not contradict the statement
in \btzpaper\ that there are no 
 closed timelike curves within the fully extended Penrose diagram.
 What happens is that 
a locally timelike curve  joining two timelike separated images 
 goes through the
so called ``singularity''. Due to this  divergence of the stress tensor 
 we can only trust the geometric description up to the inner horizon.\foot{
A related remark was made in \gary .}

These eternal black holes with angular momentum along the $\phi$ circle 
are dual to the same two copies of the CFT but now instead of the
state \wavefunction\ we have 
\eqn\wfrotating{
|\Psi \rangle = { 1 \over \sqrt{ Z(\beta,\mu)} }
\sum_n e^{ - {\beta E_n\over 2} - {\beta \mu \ell_n
\over 2}  } | E_n , \ell_n\rangle_1
\times 
| E_n , \ell_n\rangle_2
}
where $\mu$ is the chemical potential for momentum along the circle and
$\ell_n$ denotes the angular momentum of the state. 
We can do a similar construction for any other conserved charges of the 
CFT to describe charged  black holes in $AdS$, see \israel .

Given that the boundary description, in principle, describes the interior
one would  like to give a precise prescription for recovering the 
approximately local
physics in the interior, i.e. we should be able to describe approximately
scattering amplitudes measured by an observer who is behind the horizon
and falling into the singularity. 
If interactions between bulk particles are weak 
it is easy to give a prescription.  We can 
map initial and final states to states in the CFT as we
described above and then we can compute the overlap between  initial
and final states in the CFT. This can be viewed as mapping all states
to the $t=0$ slice and computing the inner product there. This is also
equivalent to computing amplitudes by 
 analytically continuing to Euclidean space in the way we explained above. 
If interactions are strong   the particles will scatter many times
before we evolve back to $t=0$. Though in principle we can do the 
computation it would be very hard in practice to extract the desired 
amplitude. 
This problem is present not  only for  black hole spacetimes 
but also  in usual  AdS/CFT. It is the problem of extracting
local bulk physics from the CFT. 
  The mapping between the 
state in the CFT and multiparticle states in the bulk can be very 
complicated, see for example \precusors . 
Certain situations, where particles get well separated from each
other after a time of order the AdS radius can be described 
as explained in \polchinski . On the other hand if we have a large number
of particles with multiple interactions within an AdS radius then it
becomes complicated to state how to recover local computations in the
interior.

 

As an aside, let us note that 
another situation with two boundaries is the case of $AdS_{d+1}$ space 
written with $AdS_{d}$ slices. The system in this case is dual to 
two field 
theories defined on 
 $AdS_{d}$ which are coupled by their boundary conditions
on the boundary of $AdS_{d}$ (see \karch ). 
 Indeed, in the bulk spacetime one can send 
signals between the two $AdS_d$ boundaries.  
This is different from the situation we
considered above, where the two field theories where decoupled. 
In our  case we do not expect to be able to send signals between the
two boundaries and indeed in the geometry we find that they separated
by a horizon.
 
If we start with a CFT with only one connected 
boundary we cannot get geometries
with two disconnected 
boundaries because they would have infinite action. When we
specify the CFT and say on which  space it lives we are implicitly 
giving a set of counterterms for the gravity solution. If we start with
only one boundary then there could exist  geometries which have additional 
boundaries but they will have infinite action and will not contribute 
to the computation if we do not include the counterterms for the extra
components of the boundary. These counterterms are local  and 
depend only on the asymptotic structure of the solution 
\refs{\wittenone,\kostas}, but we need to say over which surfaces they
are integrated. This choice of surface is the choice of space over
which the field theory is defined. 
Notice in particular that we are {\it not} interpreting \gluing\ as
a gravitational instanton giving the amplitude to create two boundaries.
The action of the euclidean  instanton is infinite, if we do not include the 
regularizing counterterms on the Euclidean boundary. Including the 
boundary conterterms is specifying precisely what theory we have, it is 
saying that we 
 have precisely the Euclidean field theory on a very specific
geometry. 
The process
depicted in \gluing\ should be thought of as the process that prepares the
entangled state, both in field theory and in gravity. 

If the curvature
of the boundary is positive and we are in  Euclidean space 
it was shown in \wittenyau\ that the boundary cannot have disconnected
pieces. If the boundary has negative curvature one can have several 
disconnected pieces. In the case of $AdS_3$ we expect to be able to 
consider the field theory on negatively curved Riemman surfaces as
long as we are not at a singular point of the CFT \swsingular . In this
situation it would be interesting to understand the meaning of
Euclidean geometries with disconnected boundaries. 

Finally let us notice that in the case of $AdS_3$ 
there are many interesting Lorentzian 
spacetimes that have multiple boundaries \multiple . These can 
be obtained by gluing suitable Euclidean geometries \krasnov .
These geometries also have the interpretation of being several
copies of the CFT in a entangled  state that can be obtained
by doing the path integral of the Euclidean theory on the boundary 
of the Euclidean geometry. In Lorentzian signature some of these
spacetimes have one boundary and some have multiple boundaries. 

It seems  that all Lorentzian spacetimes with 
multiple boundaries can be thought of as entangled states arising 
in the product Hilbert space of many decoupled conformal field 
theories \gallowaytwo  . 
In fact, it was shown  in  \gallowaytwo \gallowayone \ that in 
Lorentzian spacetimes obeying the null positive energy condition
all boundaries are screened from other boundaries by horizons. This 
is enough to ensure the vanishing of commutators for operator insertions
on different boundaries, at least to leading order in the gravity 
approximation.

\subsec{ Black holes with only one boundary\foot{
This subsection originated in conversations with G. Horowitz.}
}

One can take quotients of the eternal black hole spacetime in 
\penrose , and obtain black hole spacetimes with only one boundary. 
The main idea is to quotient by a map that acts as a  reflection
$x\to -x$ on the Kruskal coordinates \kruskalcoord\ (in other
words $u \leftrightarrow v$). This action can be accompanied by 
many other $Z_2$ actions on the full theory. 
One possibility, which was discussed in detail in \geonads , is
to also 
map a point on $S^{d-1}$  to its antipodal point, this has the
advantage of being a non-singular quotient.  
In the full string theory one might need to accompany this action
with an orientifold action, etc (see \horava\ for an  example). 
These black holes
are particular pure 
states in one copy of the conformal field theory that
lives on the boundary of $AdS_{d+1}$. By thinking about patching 
a Euclidean solution at a moment of time reversal symmetry it is 
easy to construct these states, both the state in 
 the boundary CFT and the  state for quantum  fields in the bulk.

\ifig\geon{$Z_2$ quotients of the eternal black hole. 
In (a) we see the $Z_2$ quotient of the CFT. 
It is a Euclidean cylinder going between two boundary states. 
These could be cross caps, so that we get a ``Klein bottle''. 
In (b) we cut (a) along the doted lines and view the resulting state
as the $t=0$ quantum state which is then evolved using Lorentzian
evolution.  In (c,d) we see the bulk gravitational description of
(a,b). The vertical doted line indicates that the left and right
sides are identified. In (c) the boundary cylinder of (a) is 
represented as the arc going from $\tau=0$ to $\tau=\beta/2$. 
}
{\epsfxsize3in\epsfbox{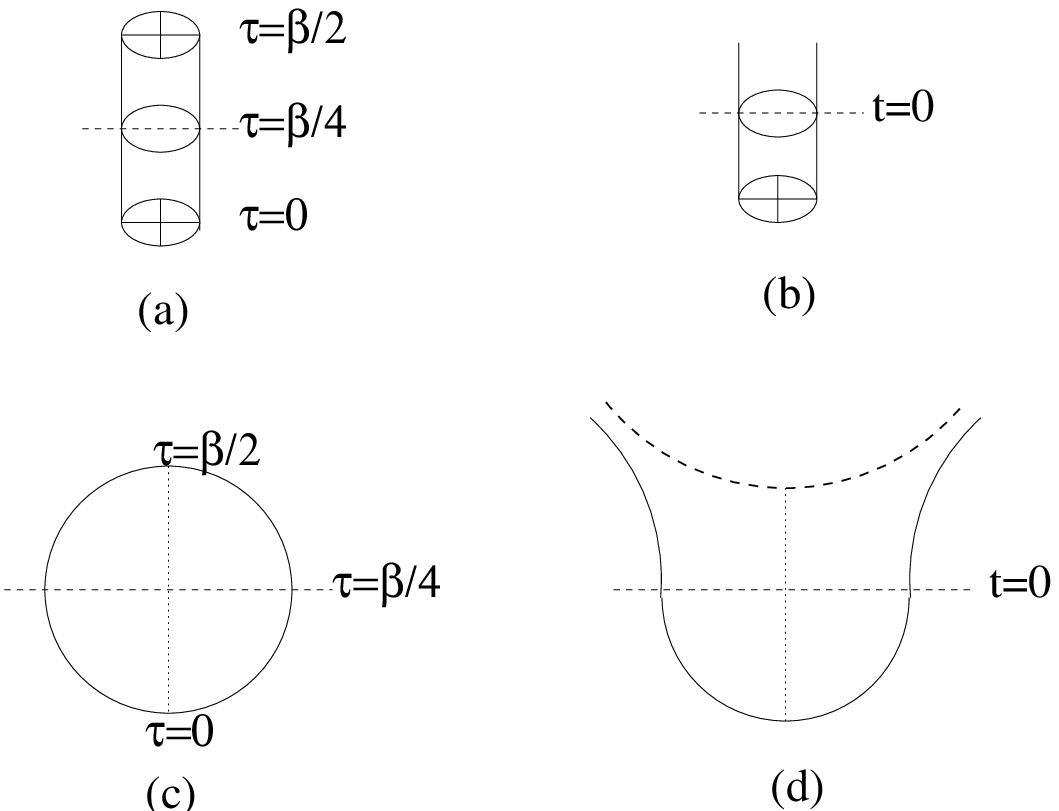}}

Let us start with the $CFT_{d}$ 
on a space which is $(S^1_\beta \times S^{d-1})/Z_2 $ where 
$Z_2$ acts by mapping a point on $S^{d-1}$ to its antipodal point
and by  a reflection  ($\tau  \to -\tau$) on   the circle $S^1_\beta$. 
The length of $S^1_\beta $ before we do the quotient 
is $\beta$. The geometry thus obtained
is non-singular, it is a ``Klein bottle'' (it is the Klein bottle
for $d=2$).
If we cut this Euclidean geometry at $ \tau = \beta/4 $, were $\tau$ is
a point on $S^1_\beta$,  we get two copies of a ``Moebius strip'' (it
is the usual Moebius strip for $d=2$). The boundary of this ``strip''
is $S^{d-1}$. So the path integral of the Euclidean theory over this
``strip'' produces a state of the CFT on $S^{d-1}$. The $d=2$ version
of this construction is very familiar,   we  produced  the 
crosscap boundary state evolved by $\beta/4$ with the closed string
Hamiltonian, which is a state in the closed string Hilbert space.
In summary, we have produced the state of the Lorentzian CFT at $t=0$
by a Euclidean path integral over a ``Moebius strip''. 

Now let us discuss  the dual  gravity description.
Let us start with the Euclidean theory. There are (at least) two
ways of filling in the Klein bottle which are related to the two
ways of filling in  $S^1_\beta \times S^{d-1}$ \wittenthermal .
 One way of 
filling it in is with a space of  topology $D^2 \times S^{d-1}$, this
is the Euclidean Schwarzschild AdS 
black hole. The $Z_2$ quotient gives 
 a non-singular space. This geometry has a  time
reflection symmetry at $\tau = \beta/4$. The spatial geometry of this
slice is that of the $t=0$ section of the quotiented Kruskal 
diagram of a black hole in $AdS_{d+1}$, so we can patch the 
Lorentzian solution to the Euclidean solution. 
This is represented pictorially
in  \geon d. As we explained above for
the eternal black hole, this 
construction determines, in a precise way, 
the quantum state for the fields in the 
Lorentzian solution. Particles in the interior are obtained by 
inserting operators in the Euclidean geometry as in \particles . 
Some aspects of this quantum state were discussed
in \geonads .
In the $AdS_3$ case 
the supergravity correlators
were computed in 
\geonads\ and are given, again, by the method of images, except that
now we have to include images under the $Z_2$ action also. This implies
that we  essentially have to  
add \same\ and \other\  together.

We can  consider other $Z_2$ quotients. For example, we 
can choose 
a $Z_2$ which purely reflects the Euclidean time
direction and does not act on the sphere. This $Z_2$ action has
fixed point on the boundary.
In fact, this   can be interpreted as a 
conformal field theory with a boundary. All we have to do is to 
substitute the crosscap in the above discussion, and in figure \geon ,
by a usual Cardy boundary state. 

In all of our discussion in this section we have assumed that 
we can actually do this $Z_2$ orbifold.  
Both in the field theory and
in string theory we have
to be careful about the presence of spinors, so that the $Z_2$ 
will also have to act  on them appropriately. 
In order to make the discussion more concrete let us mention two 
examples of $Z_2$ actions we could consider. 
As an example where the $Z_2$ action does not have fixed points
on the boundary 
consider $AdS_3 \times S^3 \times T^4$. The $Z_2$ action is an 
orientifold together with the following geometrical action 
 the Euclidean time direction, it shifts the circle $S^1_\phi$
by $\pi$ (this is the antipodal map for the circle) and it sends
$g \to g^\dagger$, where $g\in SU(2)$ describes the $S^3$.

As an example of a $Z_2$ action with fixed points on the boundary 
consider the following. 
Start from  the field theory that results from taking the low energy
limit of coincident M2 branes which is dual to $AdS_4 \times S^7$.  
Take the $Z_2$ to be a  reflection on $S^1_\beta$ as in 
\hw . This introduces a boundary in the Euclidean 
field theory, the M2 branes
can end on this boundary. The state is prepared from the boundary
state via Euclidean evolution as above. In this case the
vertical doted line in \geon\ is an end of the world ninebrane in
11 dimensions. In fact the M2 brane is stretched between an end
of the world ninebrane (at $\tau=0$) 
and an end of the world anti-ninebrane (at $\tau=\beta/2$), 
as in \hfa .
Its 
worldvolume is $I_{\beta/2} \times S^2$, where $I_{\beta/2}$ is
an interval of length $\beta/2$. 

Even though the black hole is given by a pure state we 
expect to find approximate thermal answers if we do measurements
that involve a very small subset of degrees of freedom. 
When we   only 
probe part of the system  the rest of the system is acting
as a thermal reservoir at temperature $\beta$.

\newsec{ Remarks about information loss}

The information loss paradox \hawkinginfo\ becomes particularly sharp
in AdS because we can form black holes that exist for ever. 
The information loss argument in \hawkinginfo\ says that after 
matter collapses into a  black hole all correlators with the
infalling matter decay exponentially. In \hh\ it was argued that
 computations done at late times will be the same as 
computations done in the full extended Schwarzschild geometry. These
computations show that only thermal radiation comes out and therefore
information gets lost. In the case of black holes that evaporate in finite
time there are some residual correlations with the initial state.
These  correlations are of the order of 
$e^{-c t_{evap}/\beta} \sim e^{- c' S}$ where $t_{evap}$ is 
the evaporation
time for  Schwarzschild black holes in flat space 
 and $c,c'$ are some numerical constants.  
In the case of black holes in AdS the black hole lives for ever so 
one can wait an arbitrary long time for correlations with the initial state
to decay.\foot{ For other discussions of information loss in AdS/CFT see
\lowe \jacobsonthree .} 

Here we will consider the simplest possible deformation of the perfectly
thermal Schwarzschild AdS 
state and we will show that, indeed, correlations die off
exponentially fast. 
The simplest possible deformation of the thermal state is to add an 
operator in the second boundary. From the point of view of the first
boundary this is a small change in the thermal ensemble. This change
is detectable. Indeed we can compute the one point function of 
the same operator in theory one. This correlator is zero in the perfectly
thermal ensemble (when there is no operator on boundary two), but it
is non-zero in the deformed ensemble. This non-zero value is given by
\other . In agreement with the arguments in \hawkinginfo \hh , 
this correlator decays exponentially fast as $e^{ - c  t \over \beta}$
where $c$ is a numerical constant. 
If we wait a sufficiently long time this correlator goes to zero. 
This is {\it not}  what we expect if we make a small change of the 
density matrix in a unitary theory, such as the boundary CFT. 
In some sense, we can say that the extended AdS Schwarzschild
spacetime  is ``more thermal'' than a thermal state in a 
 unitary theory.
 This is a version of the information loss paradox. It is 
a particularly sharp version of the paradox since the calculation is
very well defined.

It has been suggested that string theory, being a theory of extended
objects,  invalidates  arguments based on local field theory, such
as the arguments that lead to information loss.
It is therefore natural to ask if these effects could solve the paradox
that we have just presented. The BTZ black hole in $AdS_3$ is particularly 
useful to test this idea. We can embed $AdS_3$ in string theory in
such a way that strings moving in $AdS_3$ are described by an 
$SL(2,R)$ WZW model. Two point correlation functions in $AdS_3$ computed
using string theory have the same functional form as in field theory
\teschner \zam . 
This is easy to understand since both are restricted by conformal
invariance (i.e.  global $SL(2,R)^2$
invariance).
 The BTZ black hole is a quotient of $AdS_3$. 
The standard orbifold rules imply that the correlation function is
given by ``summing over the images''. This prescription gives  the 
stringy version of the Hartle-Hawking state. 
This produces a result that 
has the same functional form as in field theory \same \other . 
This shows that tree level string theory does not solve the problem. 
One could think that higher loops in string theory would solve the 
problem, this might be possible, but as we will see below this expectation
seems  unfounded.\foot{In the case of the SL(2,R) WZW model we find
that one loop corrections diverge as explained in \son . This is
related  to the fact that the thermal ensemble is unstable since
the black hole can evaporate by emitting long
strings. In stable thermal ensembles we do not expect such a 
 divergence. } 

Before we go on looking  for corrections we should ask:
How big are the expected correlations?
We now show that a correlation that is consistent with unitarity could
be as small as of order $e^{- c S}$ where $S$ is the entropy of the 
ensemble and $c$ is a numerical constant. Instead of presenting 
a general argument, let us do a 
calculation in a free field theory and see that 
correlations as small as $e^{-c S}$ are possible. 
The field theory we will consider is similar to 
 the so called ``long string model''
of the black hole \mss . We consider a single field $X(\tau, \sigma)$ 
that lives 
on a circle of radius  $ k$ where $k \sim 
{ R_{AdS} \over G_N^{(3)} } $, so that $\sigma = \sigma + 2 \pi k $.
The operator that we will consider is 
$ {\cal O} = \sum_n  \partial X \bar \partial X (\tau, \sigma + 2 \pi n)$.
We can use the standard formulas of finite temperature field theory \msh\
to compute the two point correlation function of an operator inserted
in the second copy with an operator inserted on the first copy. 
\eqn\resfree{
\langle {\cal O}_1(\tau,\sigma) {\cal O}_2(0,0) \rangle_{Free} \sim 
\sum_m \left[ \sum_{n=-\infty}^\infty { 1 \over
\left( 
\cosh ({2 \pi (t + 2 \pi m k )  \over \beta}) + 
\cosh ({2 \pi (\phi  + 2 \pi n) \over \beta})
\right)^{2 }  } \right] 
}
We see that the quantity in square brackets has the form of the
gravity result \other\ and the role of the sum over $m$ is to 
make the result periodic in time, under $t \to t + 2 \pi k$. 
This periodicity follows from the fact that 
all energies in this free 
theory are multiples of $1/k$. We notice, however, that between two
maxima this function is very small, it is of order 
$e^{ -  (2 \pi)^2 k/\beta } \sim e^{ - c S}$ as we wanted so show. 
In a full interacting field theory we do not expect to find a periodic
answer, but this  calculation shows us that the correlations
can be as small as $e^{- c S}$. Since the entropy is proportional to
$1/G_N$ we see that these could come from non-perturbative effects, so
there would be nothing wrong if we did not see any effect 
 in string perturbation
theory. 

Let us return to the   eternal black holes in AdS. 
The correlation functions in the boundary field theory clearly cannot decay 
to zero at large times. The problem is solved once we remember that
the AdS/CFT prescription is to some over {\it all}  geometries with 
prescribed boundary conditions. In particular, the euclidean 
thermal ensemble has other geometries besides the one we included so
far. One of them is that of an AdS space with euclidean time periodically
identified. This is a geometry with topology $S^1 \times B^{d}$. 
The fact that we should sum over geometries in the Euclidean theory
was emphasized in \refs{\hp,\wittenthermal,\ms,\dmmv}.
Since we can view the Euclidean path integral as defining our initial
wave function we will also get other geometries that contribute to 
the Lorentzian computation. The geometry that provides the effect that 
were are looking for  consists of two separate global $AdS$ spaces with
a gas of particles on them. This gas of particles is in an entangled
state. This piece of the wavefunction originates from a Euclidean geometry
which is an interval in time of length $\beta/2$ in global Euclidean AdS.
The Euclidean evolution by $\beta/2$   is responsible for creating
the entangled state for the gas of particles. 
Now if we compute the two point correlator in this geometry we indeed
get a non-decaying answer. This geometry is contributing with a 
very small weight due to its small free energy (remember the factor
$Z(\beta)^{-1/2} $
 in  \wavefunction ) compared to the AdS Schwarzschild geometry.
In fact the size of the contribution is of order 
$e^{ - \beta (F_{AdS} - F_{Black-hole} )} \sim e^{ - c' S} $ where 
$c'$ is some constant. So we get the right amplitude for the non-decaying
correlator. 

It was argued in the past that the sum over geometries would destroy
unitarity since it would involve black holes. It is amusing to note
that here  the sum over geometries  is involved in restoring 
unitarity.\foot{ G. Moore reminded me of  the connection between unitarity
in the lorentzian theory and modular invariance in the Euclidean theory. 
In the $AdS_3$ case the sum over geometries was required by 
modular invariance \refs{\ms,\dmmv}. In string perturbation 
theory this connection 
is well known, we see it appearing again in a (apparently) different 
context.} The fact that the sum over geometries  can restore unitarity
was observed  in \refs{\joeandy,\andyhouches}.

One could ask 
how to restore unitarity in the case that we start with a pure state in
a single copy of the field theory.
In fact we can consider the $Z_2$ quotients we discussed above, which 
produce black holes with a single boundary. The two point correlation
function also decays exponentially in this case. 
Once we remember that there are other ways of filling in the geometry
we realize that 
we get non-decaying contributions to the correlation function.

Finally let us remark that even though we chose to compute a
correlator between theory two and theory one, we could have computed
a correlator between two insertions in theory one. This could be 
viewed as throwing in a particle in a perfectly thermal state and asking
whether we can see any change in the state at late times. 
If we look at \same\ carefully we can also see that these correlations
decay exponentially in time (we need to convolute  with a smooth 
function at the initial point taking into account the $i\epsilon$
 prescription).

\newsec{Conclusions}

We have seen how to describe, in principle, the spacetime 
corresponding to extended AdS Schwarzschild geometries. 
More precisely, the AdS Schwarzschild geometry appears as 
the saddle point contribution of a more complicated sum over
geometries. 
This gives, in principle, the resolution of the spacelike 
singularities in the interior of black holes. 
These spacetimes are very interesting since they are simple
cosmological spacetimes. The resolution of the singularities in
this case reduces simply to the specification of initial conditions
in the full system, the full system includes other classical geometries
as well.  There are many initial conditions
corresponding to different particles coming out of the 
white hole singularity. The wavefunction $|\Psi\rangle$ \wavefunction\ 
is a very 
special choice for which we have a geometric interpretation. 
Of course these ``cosmologies'' are very unrealistic since they are
highly anisotropic (but notice that by considering black holes in 
$AdS_4$ they can be 
four dimensional).  These solutions show that it is 
possible to have cosmological looking spacetimes in non-perturbative
string theory. 
If we start the system in the state \wavefunction\ and we let it evolve,
after a time of order 
 $\beta$ we expect that the geometric description would 
break down for an important part of the state.
 But since the evolution is just adding 
phases to the state, by the quantum version of the Poincare recurrence 
theorem  after a very long time
 will  get arbitrarily close to the initial state and therefore we would
recover the geometric interpretation, but with slightly different 
initial conditions, so that we will not have the H-H state but we will have
the H-H state with some particles coming out of the white hole singularity. 
 Interpreted as a cosmology, the universe gets
to start over and over again and the initial conditions 
change slightly every time. 


It would be very nice to understand more precisely how to describe 
local processes in the interior in terms of the boundary theory.

{\bf Acknowledgements }

I would like to thank specially G. Horowitz for correcting me on one
 point. I would also like to thank 
  C. Bolech, R. Bousso, S. Giddings, P. Kraus, G. Moore, 
 A. Peet, H. Ooguri, J. Polchinski,  N. Seiberg, A. Strominger, C. Vafa,
H. Verlinde  
and all the participants of the conference Avatars of M-theory for
useful discussions and constructive  criticism.
I also thank G. Lifschytz, D. Lowe and K. Skenderis for correspondence.

I would also like to thank the ITP at UCSB  for hospitality. 
This  research 
was supported in part by DOE grant DE-FGO2-91ER40654,
NSF grants PHY-9513835 and PHY99-07949, the Sloan Foundation and the 
David and Lucile Packard Foundation.

\listrefs

\bye